# COPR – Efficient, large-scale log storage and retrieval

Julian Reichinger
Dynatrace Research
Linz, Austria
julian.reichinger@dynatrace.com

Thomas Krismayer
Dynatrace Research
Linz, Austria
thomas.krismayer@dynatrace.com

Jan Rellermeyer
Leibniz University Hannover and
Dynatrace Research
Hannover, Germany
rellermeyer@vss.uni-hannover.de
jan.rellermeyer@dynatrace.com

## ABSTRACT

Modern, large scale monitoring systems have to process and store vast amounts of log data in near real-time. At query time the systems have to find relevant logs based on the content of the log message using support structures that can scale to these amounts of data while still being efficient to use. We present our novel Compressed Probabilistic Retrieval algorithm (COPR), capable of answering *Multi-Set Multi-Membership-Queries*, that can be used as an alternative to existing indexing structures for streamed log data. In our experiments, COPR required up to 93% less storage space than the tested state-of-the-art inverted index and had up to four orders of magnitude less false-positives than the tested state-of-the-art membership sketch. Additionally, COPR achieved up to 250 times higher query throughput than the tested inverted index and up to 240 times higher query throughput than the tested membership sketch.

## 1 INTRODUCTION

Today's large scale cloud systems can already consist of tens of thousands of hardware nodes and processes. Together with the size of these systems, the complexity of detecting and understanding operational problems, security incidents or degrading user experience within them is growing continuously. As a direct response to this challenge, the monitoring of cloud systems is expanding rapidly to include additional and more fine-grained recordings of their operation, which are kept for longer retention periods to enable post-mortem analysis of past events. This leads to an explosion of machine-generated, semi-structured or unstructured data which needs to be processed, stored and analysed cost efficiently and in real-time.

Dynatrace LLC. [10] provides a data intelligence platform to its more than 3600 enterprise customers, enabling them to collect, process, retain and analyze these vast amounts of monitoring data. Dynatrace Grail [13], described as a causational data lakehouse, powers this data intelligence platform by storing and analyzing Petabytes of log, event, and metric data each day. In this paper, we introduce a new probabilistic indexing structure, named COPR, which allows Dynatrace Grail to search through up to 3.5 Terabytes of log data per second on a single CPU core (95th percentile of throughput for customer queries), while introducing a storage overhead of just over 1% of the ingested data size.

Traditional RDBMS, such as PostgreSQL [36] or Oracle Database [34], have been mainly developed to store relational data with well-defined schemas and to support transactional reads and updates [31]. However, monitoring data, like logs or metrics, is typically never updated and its attributes are dynamic and high-dimensional. NoSQL document stores, such as MongoDB [29] or CouchDB [6], achieve better horizontal scalability and relax the schema requirements of data during ingest [31]. However, the *log-structured-merge-tree* utilized by many NoSQL databases mainly supports the efficient access of data via a primary key [38]. Since the creation and maintenance of secondary indices introduces a high overhead [38], NoSQL databases often require users to define indexed data properties in advance [7, 30]. Especially the use of traditional inverted indices for full text search within these systems can produce a substantial storage overhead, as we will show in Section 5. This severely limits the usefulness for ad-hoc analysis where data access patterns are not known beforehand.

Big data processing systems, such as Hadoop Map-Reduce [16] or Apache Spark [42], try to enable complex analytics on top of big data sets by relying on highly parallel processing of raw data instead of pre-defined indices. However, a brute-force approach to data analytics cannot provide low query latencies and low analysis costs in the face of exponentially growing amounts of data, as we will also show in Section 5.

Systems aiming to support real-time processing, long-term storage and low-latency queries on big, semi-structured data sets need to combine a high degree of parallelization with novel indexing structures. These structures need to be capable of reducing the search space by several orders of magnitude while being efficient enough to support the indexing of all incoming data in real-time and without incurring a high storage overhead.

The infamous Log4Shell security incident [23] is one of the many real-world examples for use-cases requiring high data volumes, long retention times and extensive indexing. Potential attacks utilizing this vulnerability can be detected through logs, since vulnerable systems need to log some user-provided input containing the pattern "${{jndi". This pattern triggers a call to the *Java Naming and Directory Interface*, which can be exploited for arbitrary remote code executions in vulnerable systems [22]. Therefore, finding log lines with this pattern can be a strong hint toward potential attacks. However, utilizing log data for this analysis is only possible if the log data of all relevant systems was stored. Additionally, the analysis requires a search over extensive time frames, since Log4Shell existed for 9 years as a zero-day vulnerability [21, 23]. Lastly, only

very extensive indexing approaches would be able to search for a pattern of special characters.

While Dynatrace Grail [13] provides an industry-grade, distributed query execution engine, COPR is the component responsible for the efficient indexing of all customer data, and the required search-space reduction during query execution. We evaluate COPR in the context of log data retrieval. As log data is typically stored in compressed form, systems need to be able to quickly find relevant data without incurring the overhead of first decompressing the stored data [40, 47], making log data an especially important use case.

Our lead research question is, therefore, the following:

- How can queries on compressed log data be supported efficiently in terms of storage, memory and processing overhead during ingest and query execution?

We argue that our work offers several significant contributions in regards to this research question.

- We introduce a novel probabilistic indexing structure, called COPR.
- We compare COPR to state-of-the-art indexing solutions, used within the industry, and evaluate the different solutions with production and open-source data sets.
- Our benchmarks show that COPR requires up to 93% less storage compared to inverted indices, while producing up to four orders of magnitude less false positives than existing membership sketches based on Bloom filters. A log retrieval solution based on COPR is able to perform so-called needle-in-the-haystack queries up to 8 600 times faster than a linear data scan, up to 250 times faster than Apache Lucene [25] and up to 240 times faster than the CSC sketch [19].
- COPR introduces a novel algorithm for efficient, online deduplication of posting lists and postings within individual lists. This enables the usage of compressed static functions for the efficient encoding of the probabilistic indexing structure.
- We discuss several design decisions necessary to support online indexing approaches efficiently within a large-scale, distributed data lakehouse.

## 2 BACKGROUND AND RELATED WORK

A common pattern for data analysis queries is the combination of search terms, which are supposed to narrow down the data to the relevant portions, with additional mapping and aggregation steps. Furthermore, log records are typically combined into compressed batches, to utilize the redundancy between individual log lines for better compression rates.

Performing the filtering in-memory becomes prohibitively expensive for compressed log data, as it requires the decompression and scanning of potentially petabytes of data. Section 5 evaluates two implementations based on either a generic, state-of-the-art compression algorithm or a specialized compression scheme for log data, which executes searches directly on the compressed data representation. As we will show, both approaches suffer from significantly decreased query throughput, compared to solutions which avoid processing of the compressed data.

Similarly, index structures need to be capable of identifying batches matching the specified filter with high precision. Each false-positive match incurs a considerable performance penalty because of the required decompression and scanning of the entire batched record set and the associated disk IO. Section 5 includes several benchmark which show the detrimental effect of higher false-positive rates to query throughput. At the same time, the storage overhead of the index structure needs to stay at a fraction of the compressed data size. Otherwise, the disk requirements, together with the long retention times, would lead to unfeasible storage costs.

Li et. al. [19] introduced the *Multi-Set Multi-Membership-Query* (MS-MMQ) problem as a formal definition of this indexing approach. It generally describes the capability of data structures to determine which terms are included in which sets of data. E.g., which words appear within which compressed batches of log records.

Given set $S \subset \Omega$, where $\Omega$ is the universal set containing all possible elements, a *Membership Query* answers the question if for some element $t \in \Omega$, $t \in S$ also holds. Membership sketches like Bloom Filters are allowed to assume a false-positive membership with a certain probability, but must correctly determine the membership if $t \in S$ is true.

*Definition 2.1 (Multi-Set Multi-Membership Query [19]).* Given sets $S_0, S_1, ..., S_{n-1} \subset \Omega$ and element $t \in \Omega$, a *Multi-Set Multi-Membership Query* determines the membership set $M_t$ of indices for all sets among $S_0, S_1, ..., S_{n-1}$ where $t \in S_i$.

$$M_t = \{i : t \in S_i, i = 0, ..., n-1\}$$

Similar to membership sketches, MS-MM sketches are allowed to include false-positive set memberships within $M_t$ with a certain probability, but must include the membership for set $S_i$ if $t \in S_i$ is true.

Closely related to the MS-MMQ problem is the *Multi-Set Membership Query* (MS-MQ) problem, which is extensively studied, for example, in networking applications [19, 39]. However, this definition assumes that elements can only be part of a single set, which is an unrealistic assumption for our use case where terms can appear in many compressed batches. Therefore, the developed solutions cannot be directly applied to our use-cases.

Existing solutions to the MS-MMQ problem include the usage of inverted indices [25, 26], independently searchable compression dictionaries [40, 47], or probabilistic membership sketches [3, 19]. The rest of this section is dedicated to short descriptions of each of these solution classes.

### 2.1 Inverted Index

Inverted indices have been invented decades ago but are still widely used in modern database systems to quickly find data containing queried terms, effectively solving the MS-MMQ problem [11, 25, 37, 41, 43, 49]. However, inverted indices are known to require a significant amount of storage space, often exceeding the space required by the compressed log data itself. We also show this behavior in Section 5.

Inverted indices consist of a *lexicon* encoding the unique terms which appeared in the indexed data and an *inverted list* for each unique term. Inverted lists encode the sets, e.g., compressed batches, within which the term appeared. Lexicons can, e.g., be implemented as sorted lists, hash tables, or transducers [26, 27]. Since the inverted lists can grow to a significant size when large data sets are indexed, many algorithms for compact and efficient encoding and decoding of inverted lists have been developed over the last decades. Examples for list encoding algorithms are BIC [28], MILC [46], PEF [35], and others. Given a query term t, the membership set $M_t$ is equivalent to the inverted list $L_t$ of the query term.

As inverted indices are a non-probabilistic data structure, supported query patterns always produce reliable results without any false positives. Since the lexicon of an inverted index holds the full byte sequence of all unique terms, and they often include further information, such as term frequencies and term positions, they also feature extended query capabilities in addition to MS-MM queries. For example, a linear scan of the lexicon enables queries which match on parts of included terms and the stored term positions allow to search for sequences of terms appearing in a defined order.

## 2.2 Membership Sketches

Another well-studied approach to solving membership queries are probabilistic membership sketches such as Bloom Filters [3], which determine the membership of some token within a set with a small false-positive probability. They can be trivially extended for the MS-MMQ problem by keeping separate Bloom Filters for each set. However, this approach would result in problematic space and runtime complexities, as the number of Bloom Filters which have to be maintained and which need to be accessed during queries grow linearly with the number of sets. Queries would potentially need to evaluate thousands of individual bloom filters, instead of a single one.

More recent membership sketches like *Circular Shift and Coalesce* (CSC) [19] improve upon this idea by modifying Bloom Filters to directly encode the set information. Just as bloom filters, CSC uses a set of k independent hash functions $h_0, ..., h_{k-1}$ to encode the membership information of term t. The hash functions are then used to calculate k anchor positions $h_0(t)\ mod\ m$ to $h_k(t)\ mod\ m$ within a bit vector of size $m$. Set membership is stored with the help of a partitioning function $g$, which maps set identifiers to a low number of $p$ partitions. For each hash function and set $S_i$, where $t \in S_i$, the membership is encoded by setting the bit at position $((h(t)\ mod\ m) + g(S_i))\ mod\ m$ to 1. To retrieve the membership set of a query term $t_q$, the $p$ partition bits at each anchor position are retrieved and intersected. In a second step, the partitions need to be mapped to the union of sets they represent.

To reduce the false-positive rate during queries, the same process can be done for multiple *repetitions*. Each of the $j$ repetitions produces a separate sketch data structure by using an independent set of hash functions and an independent partition function. For a query with term $t$, each repetition sketch produces a separate membership set $M_t^j$. Those are then intersected to produce the final approximate membership set $\hat{M}_t$.

An arguably simpler approach has been evaluated in [48]. Each set, in this case a compressed chunk of data, is associated with a separate bit vector. Each term within the data chunk is hashed to a single position within the bit vector to encode the membership of the term.

Membership sketches usually do not provide any query capabilities beyond membership checks, but are considerably more compact than, e.g., inverted indices. A common problem is the need to define a size for the sketches before the insertion of data. Choosing an insufficient size will result in an increased false-positive rate, because most bits within the sketch will be set to 1. On the other hand, a too large size will result in an unnecessarily high space usage.

## 2.3 Searchable Compression Dictionaries

Inverted indices and membership sketches are capable of reducing the amount of compressed data chunks which need to be processed for queries. However, they necessarily produce some storage overhead, as they are always stored in addition to the compressed data. CLP [40] and LogGrep [47] mitigate this issue by employing specialized compression algorithms, which combine the search and compression capabilities into a single data structure.

The basic idea of both systems is the separation of static and variable parts of log lines. As an example, consider the log "Connection closed by 173.234.31.186" from an SSH server. While the first part is a static template, the IP address of the disconnected user varies per log line. CLP and LogGrep rely on the assumption that most logs produced by systems come from a limited number of highly repetitive, static patterns. If these static patterns can be extracted into a dictionary, each log line can be encoded with a reference to the static log pattern and the variable parts. In CLP, variables appearing at run-time are either encoded within another dictionary or directly within the data. LogGrep always encodes variables within the data, but uses a column-oriented format to improve search performance.

For both, searches are primarily performed within the dictionaries, since they are considerably smaller than the compressed data. When potential matches for search terms are found within the dictionaries, a pointer to the relevant data is required. To achieve this, both systems compress the log data in batches. Multiple batches share the same dictionaries and each dictionary entry has a list of pointers to batches containing the static pattern or variable. This structure is actually highly similar to inverted indices.

Searchable compression dictionaries, such as CLP and LogGrep, can support wildcard searches in addition to strict membership queries on tokens. However, CLP requires users to specify the static log patterns, making it unfeasible in environments where the stored logs are not known beforehand. LogGrep circumvents this issue by performing automatic pattern extraction on a small sample of the stored logs. Still, the achievable compression ratios and search performance might suffer in case of low log pattern redundancy or sub-optimal pattern extractions.

## 2.4 Self Indexes

Similar to searchable compression dictionaries, described in the previous section, self-indexes combine compression and search capabilities within a single data structure [1, 14, 15]. In contrast to CLP and LogGrep, they are not designed specifically for repetitive log data, but can operate on arbitrary text. Self-indexes compress text by encoding it as a *Compressed Suffix Tree* [14, 32]. These

Julian Reichinger, Thomas Krismayer, and Jan Rellermeyer

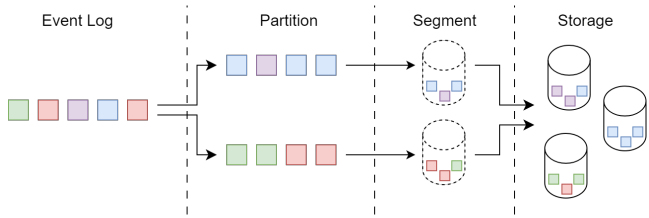

Figure 1: Top-level overview of ingest data flow

| Log message | Posting id |
|---|---|
| INFO: Connection to host established | 0 |
| INFO: Start processing | 1 |
| ERROR: Host connection terminated | 2 |
| INFO: Restart triggered | 3 |

Table 1: Example log messages

data structures can achieve compression ratios competitive to common general purpose compression algorithms for natural text [14] and also support arbitrary sub-string searches directly within the compressed text [15]. However, their search algorithms have a logarithmic dependency on the uncompressed text size [15].

Self-indexes also share a fundamental problem with most general purpose compression algorithms. They always compress a static batch of data [1]. Since data is continuously ingested, new compressed batches have to be created regularly. This fact leads to a linear run-time complexity on the number of compressed batches, because each batch needs to be checked individually for the search string. Larger batches of data will directly decrease the number of search operations, but will also lead to an increased resource usage for keeping the uncompressed data. Larger uncompressed batches can also have detrimental effects on run-time, since they can only be searched through linear data scans.

## 3 COPR APPROACH AND DESIGN

COPR has been designed to work as part of systems utilizing attribute-based partitioning of their ingest stream and organize data into eventually immutable segments, such as Dynatrace Grail [13] or Apache Druid [9]. A top-level overview of the intended ingest architecture is shown in Figure 1. Incoming records are first written to a persistent event log. Since these event logs can be re-consumed in case of errors or crashes, the following processing steps do not have to provide immediate durability guarantees after each consumed record.

The input stream from the event log is then partitioned based on some or all attributes of the individual records. This allows for a horizontal scalability of the system and can be used to group data points sharing specific properties.

In the following segmentation step, records are added to mutable segments. These segments are append-only, self-contained storage containers. They combine logic for data storage and compression, as well as data retrieval. Segments can also contain indexing structures to quickly identify query-relevant records internally. Since segments become immutable after a limited amount of data has been added to them, these indexing structures, e.g., the COPR sketch, do not have to support the indefinite addition of new data.

As these mutable segments do not have to provide any durability guarantees, internal sketches can operate completely within memory. Only once segments become immutable, the data and sketch structures need to be stored on disk. At this point, the immutable segments are stored in a replicated, distributed storage service to ensure durability and improve query parallelism.

Log records are added to mutable segments individually and are distributed to multiple, compressed batches within the segment. To efficiently locate data during queries, the content of incoming log records is first split into individual tokens. The exact tokenization strategy depends on the search requirements but one example strategy is discussed in Section 5. COPR then needs to encode and provide the information which tokens appear within which compressed batches. Section 3.2 explains the mutable, in-memory representation of the COPR sketch, which is used until segments become immutable.

### 3.1 Running example

Throughout this section we use the log messages in Table 1 to explain the different steps of the algorithm. The posting id defines which bulk of data the log message belongs to (see Section 3.2). In our example we use four fictional log lines from the same application that are stored to different postings.

### 3.2 Mutable Sketch

Within mutable segments, COPR uses a specialized, hash-based in-memory data structure, focused on combining cheap updates with instant data visibility. The basic structure is similar to a traditional inverted index, which consists of a lexicon holding the tokens and the posting lists (or inverted lists) for each token, encoding the information in which sets it appeared [26]. Each posting within a posting list identifies a single set, e.g., a batch of log records in our use-case. The mutable sketch operates on hash fingerprints within its lexicon, instead of the original tokens, and performs an online deduplication of repeated posting lists and postings within individual lists. An abstract view of the mutable sketch structure can be seen in Figure 2. Its major components are the *token map*, the *posting lists* and the *lookup map*.

We will show how each of the major components could look like, when inserting the log lines from our running example above. For this we split the complete log message into the individual words and use each (lower-case) word as one token. For example, the first log line is split into the tokens "info", "connection", "to", "host" and "established".

**Posting lists** need to encode the postings identifying all sets where a token appeared. The encoding algorithm for posting lists needs to ensure that repeated inserts of the same posting do not modify the list. Different implementation possibilities are discussed in Section 4.1. Because insertions of token-posting pairs are highly frequent operations, the focus of each in-memory posting list encoding should be efficient inserts without excessive memory usage.

The tokens from our running example only include three tokens that appear in more than one posting, "connection" (hash value 0xe3), "host" (0x32) and "info" (0x2a). "connection" and "host" are both contained in the same set of postings (0 and 2) and therefore share the same posting list (cf. Figure 2).

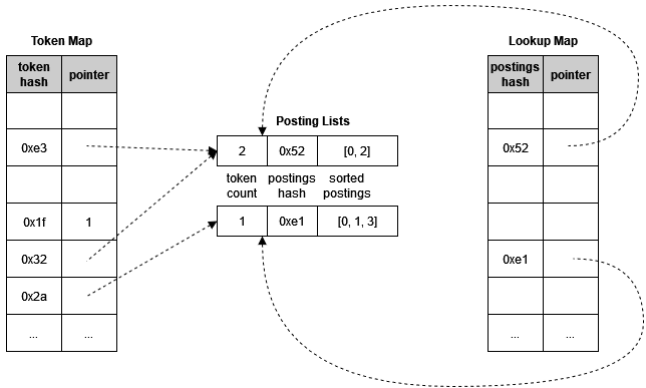

**Figure 2: Partially filled mutable sketch structure. Token hash `0x1f` references a single, directly encoded posting, while the other three token hashes encode pointers to posting lists.**

The **token map** is a hash table with fixed-sized keys and values, where the keys are hash fingerprints of the added tokens and the values identify the referenced posting list. This map serves the same purpose as the lexicon within an inverted index. Depending on the tokenization strategy for log messages, storing hash fingerprints instead of the original tokens can save a considerable amount of memory. For the tokenization strategy and the 1M_production data set introduced in Section 5, hashes require 75% less memory than full tokens. It also enables the usage of map entries with fixed sizes, instead of having to reference variable length strings from the token map. In case different tokens produce the same hash fingerprint, the token map will not be able to distinguish them and the shared posting list will contain the union of all sets the two tokens appeared in.

The token map values reference the posting list of their corresponding tokens. As one special case, the first posting of each token is directly encoded inside the value of the token map. Since literature and our own experiments showed that tokens from text documents follow Zipf's law [41], a large fraction of the tokens will only appear in a single set. Directly handling these tokens within the token map saves the computational and memory overhead of storing and managing their posting list.

Continuing the running example we can see the references for our posting lists from above. The entries for "connection" (hash value 0xe3) and "host" (0x32) reference the same posting list. The second posting list is only referenced by the token map entry for "info" (0x2a). In addition we can see one further entry for "start" (0x1f), which only appears in one posting and therefore has the posting information directly encoded in the token map entry.

During our research, we also made the observation that many tokens appear within exactly the same set of data batches, leading to duplicate posting lists. Performing a deduplication of posting lists allows us to reduce the number of posting lists by over 88% for the 1M_production data set. Especially short lists with a single or very few postings are typically shared by a high number of tokens. Having a lower number of posting lists than tokens enables the usage of *compressed static functions* for the compact representation of the references between tokens and lists in the immutable sketch, described in Section 3.3. Our algorithm includes a method for efficiently performing an online deduplication of posting lists during the iterative construction of the mutable sketch. Whenever the posting list of a token would be extended with a new posting, our method checks for an existing posting list with the required set of postings. We maintain the lookup map as a secondary data structure to perform this check efficiently.

This **lookup map** contains the hash representation of every stored posting list as a key (referred to as *postings hash*), together with the pointer to the posting list as value. Akin to the token map, the lookup map is a hash table with fixed-size keys and values. Checking for duplicate posting lists requires to check for an existing entry with the same postings hash as the new, extended set of postings within the lookup map. If a duplicate posting list is found, the value in the token map can be set to the posting list pointer stored in the lookup map value. If no existing posting list can be found for a postings hash, a new posting list with the required set of postings is stored and a new entry is added to the lookup map.

For our example there are pointers for the two posting lists described above. All other tokens only appear for one posting and therefore their references are directly encoded in the token map and not referenced by the lookup map.

Since different posting lists might produce the same postings hash, adding new entries to the lookup map needs some form of collision handling. The exact method, including the collision handling, is described in Algorithm 1 within Section 4.1. Also, whenever tokens reference a different posting list, the previous posting list might no longer be referenced by any token. In this case, their corresponding entry can be removed from the lookup map. The removal process is described in Algorithm 2 within Section 4.1.

To further speed up the deduplication process, we introduced an optimized hashing scheme. Whenever a posting list is extended, the deduplication would require the calculation of the *postings hash* over the whole set of postings included within the list. The computational effort for this hash calculation would grow linearly with the length of the list. To circumvent this issue, each posting list stores a commutative hash value over its postings as its *postings hash*. When a new posting is added to a posting list, the postings hash of the referenced list only needs to be updated with the new posting via a constant-time operation. Definition 3.1 describes the iterative calculation of the commutative hash.

*Definition 3.1.* Let $P_1$ be an arbitrary set $\{p\}$, containing a single, arbitrary posting $p \in \mathbb{N}$ and let function $hash_{element}\colon \mathbb{Z} \mapsto \mathbb{Z}$ be a uniformly distributed hash function. We define the postings hash $hash(P_1)$ of set $P_1$ as:

$$hash(P_1) := hash_{element}(p)$$

Let P be an arbitrary, non-empty set of postings and $P_1$ be the set $\{p_1\}$, where $p_1 \notin P$ and let *XOR* be the binary exclusive-or function. We define the postings hash of the union of P and $P_1$ as:

$$hash(P \cup P_1) := hash(P) \; XOR \; hash(P_1)$$

Since the binary XOR function is commutative by definition and the hashes of the elements are independent of each other, the postings hash function is also commutative. We resort to *linear congruential generators (LCG)* for the $hash_{element}$ function for single postings, as they are computationally cheap and produce approximately uniformly distributed, pseudo-random sequences of integers with a full period [44] (see Definition 3.2).

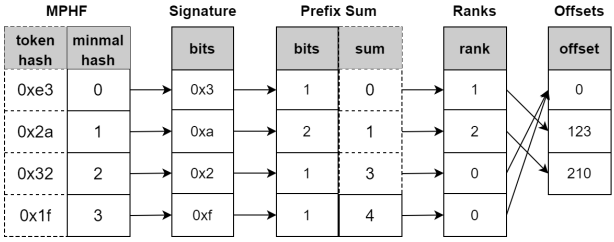

Figure 3: Immutable sketch structure created from the mutable sketch. It utilizes a compressed static function to encode the references between tokens and the deduplicated posting lists. Dotted entries are only included for readability and are not actually stored. Prefix sums are, e.g., only stored for some of the entries.

*Definition 3.2 (Linear Congruential Pseudorandom Number Generator (LCG) [44]).* A *linear congruential pseudorandom number generator* is defined as a recurrence of the form

$$x_n = (ax_{n-1} + c) \bmod m$$

where $m \in \mathbb{Z}$ is the *modolus*, $a \in \mathbb{Z} \cap [1..m)$ is the multiplier, $c \in \mathbb{Z} \cap [1..m)$ is a non-zero constant and $x_n \in \mathbb{Z} \cap [0..m)$ is the state of the generator after step $n$.

To compute the posting hash for a single posting $p$ we set $x_0 = p$ in the LCG and use the resulting $x_1$ as $hash_{element}(p)$.

For example, to calculate the posting hash for the posting list for postings 0, 1 and 3 for the token "info" from our running example we have to combine the hash values for the three postings. If the three postings give the hash values 0xad, 0x61 and 0x2d, the postings hash is calculated as (0xad *XOR* 0x61) *XOR* 0x2d = 0xe1.

### 3.3 Immutable Sketch

Once a segment becomes immutable, the focus of the sketch needs to shift from cheap updates to a small storage overhead. At the same time, the fast access times for queries and low false-positive rate must be retained. Figure 3 shows an abstract view of the immutable sketch.

While the *token map* could be stored directly, each entry requires several bytes for the key and value. Furthermore, the space requirements will be significantly increased depending on the load factor of the hash map. *Minimal Perfect Hash Functions (MPHF)* are a well studied class of algorithms for the encoding of dictionaries [2, 20, 32] and have also been considered for the encoding of the lexicon in inverted indices [26]. For a static set $S$ of keys, a minimal perfect hash function is an injective function which maps each key $s \in S$ to exactly one value within the range $[0, |S| - 1]$. In a first step, we build an MPHF for all keys inside the token map and then associate the perfect hash values with the corresponding token map values. The theoretical lower bound on the required space of an MPHF is roughly 1.44 bits per key [2] and there exist practical implementations which achieve less than 3 bits per key [20]. This approach would already remove the need to store the token fingerprints and alleviates the problem of the hash map load factor. Even though MPHFs typically only operate on static sets of keys [20, 26], their usage becomes feasible due to the eventual immutability of COPR.

When a MPHF is used to map a key to an associated value, it is referred to as a *static function* [2]. In our case, the values associated with each token fingerprint are the posting list identifiers. At this point, all single-value posting lists, directly encoded as a token map value, need to be encoded as full posting lists. This ensures that all entries in the token map reference a deduplicated posting list. Due to the deduplication, we expect the number of posting lists to be significantly smaller than the number of tokens. Additionally, some of these posting lists will typically be referenced by a larger number of tokens than others. Whenever the distribution of values is skewed, entropy based encoding can be used to compress the value representation in static functions [2].

To get the compressed value representation, the set $P$ of posting lists is first sorted by the number of tokens referencing each list. The posting list with most references will then be assigned *rank* 0, while the posting list with the least references will receive *rank* $|P|-1$. The order of posting lists with an equal number of references makes no difference. While previous work on *Compressed Static Functions* used Huffman codes to encode these ranks [2], the unique decodability of values is actually not necessary, as we will show later. Instead, we use $\lfloor log_2(max(rank_p, 1)) \rfloor + 1$ bits to encode the binary representation of the rank of any posting list $p$.

For each minimal hash value, starting at 0, we obtain the rank of the corresponding posting list and append the encoded binary representation to a bit sequence. As this produces a variable length encoding of the posting list ranks, an additional prefix sum array [2, 32] is needed to find the correct offsets within the bit sequence for each minimal hash value. Since the prefix sum array stores the bit length of each entry, we can use this information to correctly decode the rank, without storing uniquely decodable values. The prefix sum also ensures constant access times, as it stores a total offset into the bit sequence at a configurable interval. To finally locate a posting list on disk, an additional mapping from the rank to its offset is required. However, since the number of posting lists is considerably smaller than the number of tokens, this only adds a negligible storage overhead.

MPHFs can have a theoretical lower bound of 1.44 bits per key because they are allowed to produce an arbitrary mapping for keys not in the initial set $S$ of keys [2, 12]. Because of this, querying a token which was never added to COPR would always lead to false-positive matches, as it would be treated as an arbitrary token from the initial key set. The false-positive rate can be reduced by storing a configurable number of signature bits for every token in the initial set [2]. When the minimal hash value of a token is accessed, the corresponding signature bits can be compared with the queried token. A signature of $b$ bits reduces the false-positive rate for tokens outside the initial set by a factor of $2^b$.

The encoding of posting lists is an intensely studied field with a wide range of available algorithms [28, 35, 46]. Any storage efficient encoding could be used within COPR. Section 4.2 contains more details on our chosen algorithm and the reasoning behind its usage.

## 4 IMPLEMENTATION

The previous section described the main concepts of COPR's mutable and immutable membership sketches. As our approach offers a lot of flexibility for the concrete implementations of individual parts, this section will be concerned with the specific algorithms used by our reference implementation.

### 4.1 Mutable Sketch

Within the *token map*, we use 4 byte hash fingerprints for the tokens. The decision on the hash size is mainly a compromise between acceptable false-positive rate, memory usage and alignment to memory words. As we typically expect a few million unique tokens within a single segment, $2^{32}$ different hash values offer a good compromise within Dynatrace Grail. However, other use-cases could potentially benefit of using 2-byte or 8-byte hashes. The values for each token map entry are also encoded within 4 bytes. We use the two most significant bits within each value to distinguish between absent values, directly encoded postings and posting list pointers. This leaves us with $2^{30}$ possible postings and posting lists, easily enough for even our biggest expected data sets handled by a single COPR sketch.

COPR currently distinguishes between short and long lists for the *posting list* encoding. Short lists, with posting counts below a configurable threshold, are encoded as sorted lists of 2 byte positive integer values. Maintaining a sorted set of postings, with a fixed-length encoding, enables the use of binary search to check for the presence of postings when new token-posting tuples are added. Longer lists are represented as dense bitsets. The usage of bitsets and the limit for postings within sorted lists effectively leads to a constant time complexity for inserts. As a trade-off, users need to limit the number of postings to at most $2^{16}$. A smaller number of postings directly decreases the memory usage and disk usage of posting lists, because less postings need to be encoded. However, it also increases the amount of data which needs to be accessed for queries. If a user configures, e.g., 1024 postings, any query matching a single data batch needs to read at least a 1024th of the data. While this encoding strategy fitted our tested use-cases and parameters well, it could easily be exchanged, depending on the expected workload.

Whenever a posting list is extended, a deduplication is performed. This ensures that all tokens with the same set of postings reference the same posting list. Algorithm 1 describes the deduplication process of posting list entries within the *lookup map*. In most cases, posting lists are added to the lookup map with their postings hash as the lookup key. However, if two or more distinct posting lists have the same postings hash, newly added posting lists are inserted at the next highest, unoccupied hash.

Each posting list additionally holds a 4 byte *token count* field, keeping count of the tokens which reference the posting list. This reference counting allows us to safely deallocate a posting list once no tokens reference it anymore. Once a posting list can be deallocated, its corresponding entry also has to be removed from the lookup map, again requiring a form of collision handling compatible

---

**Algorithm 1** Insertion of a posting list into the lookup map

**Require:** posting list **P**, lookup map **L**
  $h \leftarrow hash(P)$   ▷ Definition 3.1
  **while** $h \in L$ **do**   ▷ Skip colliding entries
    $P_c \leftarrow L[h]$   ▷ Acquire posting list with hash value
    **if** $P \equiv P_c$ **then**   ▷ Check postings equality
      increase token count of $P_c$   ▷ Track references to list
      **return**
    **else**
      $h \leftarrow h + 1$   ▷ Hash collision found
    **end if**
  **end while**
  L[h] = P   ▷ Store reference in lookup map

---

**Algorithm 2** Removal of a posting list from the lookup map

**Require:** posting list **P**, lookup map **L**
  $h \leftarrow hash(P)$
  **while** $h \in L$ **do**   ▷ Find correct entry
    $P_c \leftarrow L[h]$   ▷ Acquire posting list with hash value
    **if** $P \equiv P_c$ **then**   ▷ Check postings equality
      remove key h from L   ▷ Identified correct entry
      **break**
    **else**
      $h \leftarrow h + 1$   ▷ Hash collision found
    **end if**
  **end while**
  $h_f \leftarrow h$   ▷ Freed entry
  $h \leftarrow h + 1$   ▷ First entry which might be moved
  **while** $h \in L$ **do**   ▷ Check all lists until the next free entry and try to move them closer to their original hash
    $P_c \leftarrow L[h]$
    $h_c \leftarrow hash(P_c)$   ▷ Intended entry for list
    **if** $h_c \leq h_f$ **then**   ▷ Needs to be moved
      remove key h from L   ▷ Remove from current entry
      $L[h_f] = P_c$   ▷ Insert at free entry
      $h_f = h$   ▷ New entry is freed
    **end if**
    $h \leftarrow h + 1$
  **end while**

---

to the strategy used in Algorithm 1. The removal process including the collision handling is described in Algorithm 2. Starting at the postings hash of the list, the algorithm searches through all entries with an equal or higher hash value until the list is found or an empty entry is reached. Once the list is removed, its entry becomes unoccupied. Postings lists which previously collided with it, can then be moved closer to their actual postings hash. This ensures that the algorithms for insertion and removal can always stop their searches once an unoccupied entry is reached. In practice, numeric overflows of hash values have be handled within the removal process. We omitted this handling from the pseudo code in Algorithm 2 for simplicity and space reasons.

## 4.2 Immutable Sketch

*Minimal Perfect Hash Functions* can be constructed with algorithms based on hypergraph peeling [4], Gaussian elimination [12] or bit-vector hashing [20]. We chose the bit-vector based *BBHash* algorithm for our immutable sketch, as it offers drastically improved construction speeds compared to the other state-of-the-art approaches [20]. Even though other algorithms produce MPHFs closer to the theoretical minimum of 1.44 bits per key [20], the slightly increased space usage does not outweigh the performance gains for our use cases. In [20], the smallest competing algorithm requires only 30% less space than the suggested configuration of *BBHash* but takes 5 times longer to construct. Additionally, the construction requires 40 times more memory than *BBHash*.

For the encoding of the posting lists, we resort to *Binary Interpolative Coding (BIC)* [28]. Even though many newer algorithms have been developed since its introduction, it still offers the most succinct representation of posting lists across all state-of-the-art algorithms [35]. BIC is a bit-aligned encoding and therefore avoids the necessary overhead of all byte aligned encodings. Additionally, BIC very naturally handles clusters of postings and can achieve an average encoding size of less than 1 bit per posting [28, 35]. As a drawback, BIC offers significantly slower decoding speeds than many newer algorithms [35]. However, we argue that the decreased decoding speed is no practical disadvantage in our scenarios. Since the MPHF and the signature bits are sufficient to check the presence of a token within the sketch, posting lists are usually not decoded at all for needle-in-the-haystack queries. For tokens which are included within the sketch, each decoded posting will trigger the decompression and post-filtering of a storage batch containing several hundred kilobytes of data. This will significantly outweigh the 5 nanosecond decompression time per posting of BIC [35].

Queries over large enough time-frames will need to query a high number of individual sketches and it will not be possible to keep all required sketches permanently in memory. It is therefore necessary to keep the overhead of preparing a stored sketch for query execution well below the cost of the actual query. Otherwise it might become the dominant factor for queries over large amounts of data. To achieve this goal, it is necessary to avoid any deserialization of the immutable sketch and any construction of in-memory data structures, as far as possible, when opening a reader.

The *BBHash* algorithm is based on bit vectors which are stored sequentially and queries directly operate on these bit vectors [20]. Also the signature bits and prefix sum arrays can be stored as sequential bit sequences and can be directly used without further deserialization [32]. *BIC* encoded posting lists only require the initial value range of posting IDs to decode the bit aligned representations again [28]. Since we store the offsets of each unique posting list within a prefix sum array, all posting lists can be sequentially encoded to a single bit sequence. Overall, only a few dozen bytes of configuration and run-time information are needed to determine the file layout of the immutable sketch. No other data needs to be loaded and parsed when opening a reader for the immutable sketch, everything else is only required during query execution.

Loading the sketch into memory might become another performance bottleneck for query execution. Querying a token which is not included in the sketch, only requires accesses to a few positions within the *MPHF* and a single position within the signature bits. However, when not all sketches can be kept in memory, queries might require to load a sketch with several MB into memory. To avoid this, each immutable sketch operates on a single memory mapped file. The operating system will ensure that only the accessed disk pages are loaded into memory and only the most frequently used sketch parts remain in memory. To further minimize the number of disk pages which need to be accessed for query execution, we collect all data required for reader opening in a header section at the beginning of the encoded sketch. This way, opening a reader will typically only need to access a single disk page.

## 4.3 Segmentation

An important practical consideration when employing sketches like COPR in large-scale data systems, is their memory usage. Since many segments can be constructed in parallel, the memory usage of each segment and sketch must only use a small fraction of the total system memory. In order to guarantee this, we implemented an internal segmentation process within COPR. Whenever the estimated memory usage of the mutable sketch structure exceeds a configured maximum, a temporary immutable sketch is constructed and written to disk. As explained in Section 4.2, the immutable sketch structure can be mapped into memory. This way, the operating system will only keep those parts of the COPR segments in memory which are actually used and do not exceed the physically available memory of the system.

Accessing the postings of a token requires to access each segment individually and merging the postings. Aside from this query overhead, individual segments typically also require more disk space than a single immutable sketch. Therefore, we want to merge the segments into a single sketch once COPR becomes immutable. Since MPHFs do not contain the original keys, merging them directly is not possible. To circumvent this issue, we store the full token fingerprint instead of a few signature bits for temporary segments. This way, it is possible to iterate over all minimal hashes and retrieve the corresponding token fingerprints and posting lists. From each segment, all unique posting lists and their referencing token fingerprints are then added to a single mutable sketch. At the end this mutable sketch contains all information, as if no internal segmentation had been used. Of course, this will necessarily also lead to the same memory usage as if no segmentation was used. This is no issue in practice, as the transformation from mutable to immutable segments can be an asynchronous operation and only a small number of segments will be transformed at any point in time.

## 4.4 Query Execution

Queries follow the same general process for mutable and immutable sketches and can be evaluated on both types. Only the data structures from which certain points of information are retrieved differ between them. Each query consists of a list of tokens for which the corresponding posting lists should be retrieved and a consumer for the decoded posting lists. Algorithm 3 shows the abstract execution logic of a query.

First, the query tokens are hashed with the same hash function used during ingest to produce the token fingerprints. Next, for each query token present in the sketch, we acquire an ID uniquely

COPR – Efficient, large-scale log storage and retrieval

**Algorithm 3** Query Execution

**Require:** query tokens **T**, postings consumer $c$
    $listIds \leftarrow \{\}$   ▷ Set of posting list IDs
    **for** $t \in \mathbf{T}$ **do**
        $h \leftarrow hash(t)$   ▷ Hash the query token
        **if** $isPresent(h)$ **then**
            $listIds \leftarrow listIds \cap acquireList(h)$
        **else**
            $c.accept([\,])$   ▷ Notify about empty posting list
            **if** $c.shouldStop()$ **then**   ▷ Abort if requested
                **return**
            **end if**
        **end if**
    **end for**
    **for** $id \in listIds$ **do**
        $l \leftarrow decode(id)$   ▷ Decode unique posting list
        $c.accept(l)$   ▷ Pass decoded postings
        **if** $c.shouldStop()$ **then**   ▷ Abort if requested
            **return**
        **end if**
    **end for**

identifying the referenced posting list. Within the mutable sketch, the method $isPresent(h)$ can look up the token within the token hash map to check the presence. For $acquireList(h)$, the mutable sketch needs to associate each unique posting list with a unique identifier internally. The immutable sketch needs to calculate the minimal perfect hash for the query token and compare the stored signature bits to perform the $isPresent(h)$ check. Since each unique posting list is only stored once, the storage offset of each posting list can be used as the unique identifier in $acquireList(h)$.

For tokens not present in the sketch, the consumer is informed about the empty posting list. This is necessary in case the consumer combines the individual results of each token via Boolean logic, e.g., to find chunks of data where all queried tokens appear. For each unique posting list identifier, we decode the postings of the list and pass the decoded postings to the consumer. Since we expect the number of unique posting lists to be significantly smaller than the amount of unique tokens, some query tokens might reference the same posting list. Passing the same posting list to the consumer multiple times has no functional use, but would result in unnecessary decoding overhead. We avoid this by decoding each unique posting list only once.

After each passed posting list, the consumer can indicate to terminate the query execution early. This is especially useful for Boolean queries which cannot produce any further matches after evaluating parts of the query tokens.

## 5 BENCHMARK EVALUATION

All experiments have been performed on an AWS EC2 *i4i.4xlarge* instance running the Amazon Linux 2023 operating system. This instance type is optimized for storage-intensive applications like database systems and offers dedicated NVMe-SSDs [18]. Our benchmark suite has been developed in Java and all experiments were executed on OpenJDK 17.0.7 [33]. All experiments were executed with warm-up iterations and multiple, averaged measurement iterations for singe-shot benchmarks.

Each tested implementation needs to be capable of storing log data and retrieving all matching log lines for queries. We encoded these requirements as a common log store interface, which each implementation needs to fulfill. During ingest, log lines are passed individually to the log store. After the whole test data set has been ingested, the log store will become immutable. At this point, all data needs to be synced to disk and the final representations of indices or sketches must be constructed. This setup closely matches the segment structure explained in Section 3. Queries can either search for full terms or perform wildcard queries, where the queried term can be an arbitrary sub-string within the log line.

All log stores, except the LogGrep implementation, collect the log lines into a configurable number of batches and use the zStandard algorithm [50] for data compression. Indices or sketches can then be used to locate batches which might contain relevant data for a query. To only retrieve matching log lines, the Boyer-Moore algorithm [5] is used to post-filter all log lines within a located batch. Performing data decompression and post-filtering during query benchmarks shows the realistic impact of false positives on query performance. This allows us to fairly compare cheap and fast sketches with slower, but more accurate approaches like inverted indices.

We evaluated the following implementations regarding ingest throughput, disk usage and query performance. To utilize the common benchmark framework, all log stores are implemented in Java.

- **COPR**: Indexes and retrieves relevant batches with the COPR membership sketch.
- **Lucene** [25]: Indexes and retrieves relevant batches with Apache Lucene, representing an industry-proven, state-of-the-art inverted index implementation. We used version 9.6 for our experiments. Since our benchmark scenarios only query individual tokens, no token positions or term frequencies are indexed, as they would only increase the disk usage without improving the query performance.
- **CSC** [19]: Indexes and retrieves relevant batches with CSC, representing a state-of-the-art membership sketch implementation.
- **LogGrep** [24]: LogGrep internally handles compression and search, and directly evaluates queries on the compressed data. It represents a state-of-the-art system for building a searchable, compressed representation of repetitive log data. Since the LogGrep implementation has an internal limit on the data set size, we create batches of at most 128000 log lines and search each batch individually during queries. We implemented a Java wrapper around the C++ reference implementation for the integration into the common benchmark framework.
- **Scan**: Implementation of a brute-force search solution that does not build any supporting data structures for queries and solely relies on post-filtering. This is used as a baseline for the experiments and represents the performance of a system which needs to decompress and post-filter all data during a query execution.

All benchmarks were performed single-threaded. We focused on the single-thread performance of the different implementations,

| Data | Lines | Sources | Size [MB] |
|---|---|---|---|
| 1M_production | 1 198 175 | 3 234 | 380 |
| 5M_production | 5 499 095 | 60 580 | 2 160 |
| 1M_generated | 1 046 661 | 3 234 | 139 |
| 5M_generated | 7 033 608 | 60 580 | 1 177 |

Table 2: Summary of production and synthetic data sets used for our evaluations.

because horizontal scaling in segment-based, distributed storage solutions can be achieved through data partitioning.

We evaluate the log stores against four different data sets. Two of them are production data sets of different sizes, consisting of log data produced by the self-monitoring of our Dynatrace Grail production clusters. Since our clusters consist of a multitude of different services, these data sets are highly heterogeneous. However, monitoring solutions like Dynatrace typically know the source of a log line, e.g., the service instance which produced it. This information is available within the data sets as an abstract source identifier. Log stores can use this information to group log lines from the same source together to improve the compression ratios and query speed.

Since the production data sets contain confidential data, we are not able to make them publicly available. To ensure the reproducibility of our results, we provide a data set generation tool together with our benchmark framework. This tool can generate data sets based on the publicly available LogHub collection [17], which closely match the statistical properties of our production data. To achieve this, we recorded the distribution of log lines per source identifier in the production data. Our data generation tool can use these distributions to generate matching data sets. The production data distributions are available within the benchmark framework. Additionally, we will show in the evaluations that all log stores show similar performance characteristics over the generated and the production data sets. Table 2 contains a summary of all data sets used in the evaluation. The data set sizes between one and seven million log lines cover the realistic size range we would expect for a single segment in production systems.

## 5.1 Ingest and disk usage

All sketches and indices necessarily incur some amount of CPU and storage overhead, since they need to be constructed and stored together with the data. In order to warrant their usage, the benefits they bring during query execution must outweigh this initial overhead. Since log data often needs to be retained for several years, we consider the storage overhead as the main cost factor of indexing structures. We will start the evaluation by comparing the overheads and overall ingest and storage costs of each log store implementation. All ingest experiments were performed with three warm-up iterations and three measurement iterations. The average over all measurement iterations is reported as the final result.

*5.1.1 Ingest configuration.* Since the tokenization strategy directly influences the processing and storage overheads for sketches and indices, we will first explain which types of tokens are produced from log lines. Tokens are formed according to the following rules.

(1) Sequences of alphanumeric ASCII characters
(2) Sequences of non-alphanumeric ASCII characters (e.g., "\${{")
(3) Sequences of non-ASCII characters (e.g., "äöü")
(4) Sequences of two alphanumeric tokens separated by a single separator character ([.:-_/@]; e.g., "name@company")
(5) Sequences of three alphanumeric tokens separated by single dot characters (e.g., "192.0.0")
(6) Each alphanumeric ASCII token is split into all included 3-grams (e.g., "warning" is split into "war", "arn", "rni", "nin" and "ing")
(7) Each non-alphanumeric ASCII token is split into all included 1-grams, 2-grams and 3-grams (e.g., "\${{" is split into "\$", "{", "{", "\${" and "{{")
(8) Each non-ASCII token is split into all included 2-grams (e.g., "äöü" is split into "äö" and "öü")

For the CSC and COPR log stores, all listed tokenization rules are used. The n-grams produced by rules 6–8 allow these stores to perform almost arbitrary contains queries, even though none of them store any original tokens within the sketch. Since Apache Lucene keeps the original tokens within its dictionary, it can perform a scan through this dictionary to perform contains queries. Therefore, the Lucene log store only utilizes tokenization rules 1–5.

Besides the tokenization logic, the allowed memory usage is another performance-relevant factor for COPR and Lucene. Both implementations allow to flush intermediate segments to disk when some memory threshold is reached. Generally, the overhead for the final merge of the intermediate segments grows with the number of these segments. Therefore, a higher memory threshold will typically improve the ingest performance. To ensure a fair comparison, both implementations are allowed to use 32 MB of memory in our experiments and merge all intermediate segments into a single, final segment when the benchmark is finished.

The CSC sketch does not support any internal segmentation, since individual sketches cannot be merged afterwards. Therefore, its memory usage is necessarily equal to its disk usage. CSC has been configured to use a single repetition with four hash functions. This configuration achieves roughly the same ingest speeds as our COPR sketch (see below). Utilizing additional hash functions or repetitions would severely impact the ingest performance.

*5.1.2 Ingest speed.* Figure 4 shows a comparison of the ingest speeds over all data set and log store combinations. To ensure a fair comparison between different approaches, e.g., sketches and searchable compression dictionaries, all indexing or processing steps are included within the measurements. For sketch and index based stores, the *ingest* time includes tokenization and indexing, in addition to log storage and compression. The *sketch_finish* time describes the time needed to build the final sketch or index representations and write them to disk. Finally, the *data_finish* time holds the required time for flushing and compressing all log lines which were still buffered in memory when the ingest benchmark was completed.

We can see that the COPR, CSC and Lucene stores have comparable ingest times, even though the sketch-based stores need to process up to 3.3 times more tokens. COPR even manages to outperform Lucene for two out of four data sets. If arbitrary *contains queries* are not an important use-case, tokenization rules 6 to 8 could be disabled for the sketch based approaches. This would further lower the ingest time for COPR by 43%–60%. LogGrep consistently performs worst over all tested data sets. This is especially interesting in comparison to the Scan store, since both, in essence, only need to store and compress the log data. However, LogGrep's compression approach seems to be considerably more processing-intensive. Figure 4 shows that the LogGrep compression even exceeds the total ingest time needed for all sketch and index based approaches, over all data sets.

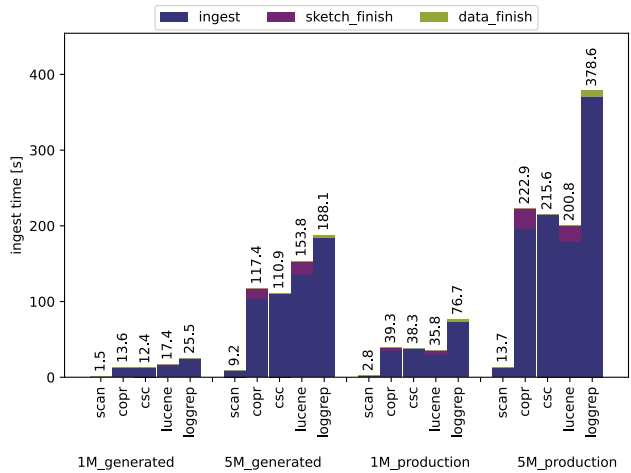

Figure 4: Comparison of ingest speeds over all log stores and data sets (lower is better).

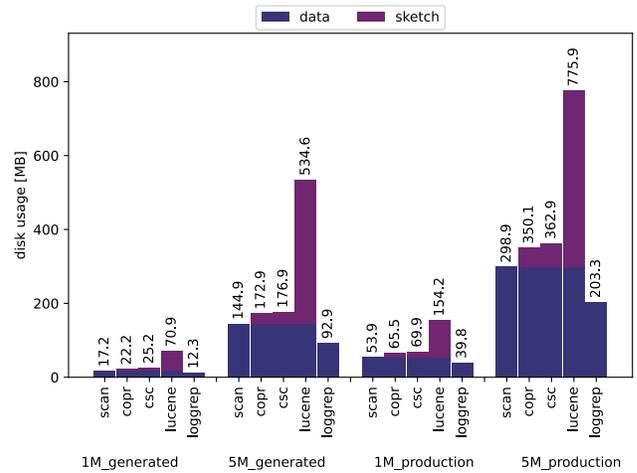

Figure 5: Comparison of disk usage over all log stores and data sets (lower is better).

*5.1.3 Disk usage.* Figure 5 shows the disk usage of each data set and log store combination. We further split the measurement into the *data* part, representing the compressed log lines, and the *sketch* part, representing the size of the sketch or index structures. We can see that the size of the Lucene index is consistently greater than the compressed log data itself. The overhead ranges between 160% and 312% when compared to the size of the compressed data, and between 12% and 39% when compared to the raw data size. Within a production system, this would directly translate to an increase of the storage costs by at least a factor of 2.6. In comparison, the overhead of the COPR store ranges between 17% and 29% compared to the size of the compressed data, and between 2.3% and 3.6% when compared to the raw data size. This is roughly 90% less overhead than Lucene.

For the CSC sketch, the disk size is directly controlled via the configuration. We decided to always size it at the next power of two higher than the size of the COPR sketch. Power-of-two sizes enable our CSC implementation to perform necessary modulo calculations via efficient bit operations. We will show in Section 5.2 that the CSC sketch still leads to considerably higher error rates for some queries, even though we assigned it a larger size than our COPR sketch. LogGrep manages to achieve up to 36% smaller data sizes than the Scan store, which utilizes the zStandard compression.

## 5.2 Queries

After investigating the overhead introduced by sketches and indices, we will now take a look at the benefits they provide for query executions. Table 3 contains a summary of the achieved query throughput for different data store implementations, data sets and query scenarios. All queries in the table have been performed in a cold query mode, where queries need to open a new reader for each execution and the page cache of the operating system is cleared between executions. This forces indices and sketches to actually perform IO operations and closely simulates expensive queries on large amounts of data, where even the sketches or indices exceed the physical memory. We primarily focus on this query mode, since it represents the worst-case scenario for query throughput. To enable a fair comparison between all approaches, queries need to decompress data batches identified by indexing structures and post-filter all contained log lines to only deliver matching results. This way, a higher number of false positive matches from sketches will result in a decreased query throughput. All query experiments were performed with a one minute warm-up iteration and a one-minute measurement iteration.

We deliberately focus on comparatively simple, but still realistic and relevant, queries, since they allow us to showcase the strong and weak points of each evaluated solution in a comprehensible way. In more complex query scenarios, overlapping effects make it hard to explain the achievable performance. Therefore, we argue that simpler query scenarios actually allow us to provide deeper insights into the individual approaches, through better separation of concerns.

First, we will examine needle-in-the-haystack queries, which search for terms that do not appear within the vast majority of stored data. This is the use-case where indices and sketches can offer the biggest benefit over linear data scans. Our *ID* queries search for randomly generated sequences of 16 English letters (e.g., "lamhmhiagialitjl"). They represent needle-in-the-haystack queries for unique identifiers of, e.g., users, processes, or devices. *Term* queries only need to find log lines where the whole query term was indexed as a separate token. In contrast, *contains* queries need to find all occurrences of the query term within or across token borders.

| Query | Data | Scan | COPR | CSC | Lucene | LogGrep |
|---|---|---|---|---|---|---|
| term(ID) | 1M_generated | $6.4e+0$ | $7.7e+3^*$ | $4.3e+3$ | $3.9e+3$ | $1.6e+0$ |
| | 1M_production | $2.4e+0$ | $4.6e+3^*$ | $2.9e+3$ | $2.9e+3$ | - |
| | 5M_generated | $8.1e-1$ | $4.2e+3^*$ | $2.7e+3$ | $2.5e+3$ | $2.2e-1$ |
| | 5M_production | $4.4e-1$ | $3.3e+3^*$ | $1.9e+3$ | $2.0e+3$ | - |
| contains(ID) | 1M_generated | $6.4e+0$ | $5.5e+3^*$ | $4.2e+3$ | $3.3e+1$ | $1.6e+0$ |
| | 1M_production | $2.4e+0$ | $1.7e+3^*$ | $1.8e+3^*$ | $9.8e+0$ | - |
| | 5M_generated | $8.1e-1$ | $2.8e+3^*$ | $2.7e+3^*$ | $1.1e+1$ | $2.2e-1$ |
| | 5M_production | $4.4e-1$ | $3.5e+2^*$ | $3.6e+2^*$ | $2.6e+0$ | - |
| term(IP) | 1M_generated | $5.1e+0$ | $7.7e+3^*$ | $2.4e+3$ | $3.9e+3$ | $1.5e+0$ |
| | 1M_production | $2.0e+0$ | $4.6e+3^*$ | $1.8e+2$ | $2.9e+3$ | - |
| | 5M_generated | $6.5e-1$ | $4.3e+3^*$ | $2.7e+1$ | $2.5e+3$ | $2.1e-1$ |
| | 5M_production | $3.7e-1$ | $3.2e+3^*$ | $1.3e+1$ | $2.1e+3$ | - |
| contains(IP) | 1M_generated | $5.1e+0$ | $6.2e+0$ | $6.2e+0$ | $3.3e+1^*$ | $1.5e+0$ |
| | 1M_production | $2.0e+0$ | $2.1e+0$ | $2.1e+0$ | $9.8e+0^*$ | - |
| | 5M_generated | $6.5e-1$ | $6.8e-1$ | $6.8e-1$ | $1.1e+1^*$ | $2.1e-1$ |
| | 5M_production | $3.7e-1$ | $3.8e-1$ | $3.7e-1$ | $2.5e+0^*$ | - |
| term(extracted) | 1M_generated | $5.3e+0$ | $2.1e+2^*$ | $2.0e+2^*$ | $1.8e+2$ | - |
| | 1M_production | $2.2e+0$ | $4.0e+1^*$ | $3.7e+1$ | $3.8e+1^*$ | - |
| | 5M_generated | $6.6e-1$ | $1.2e+1^*$ | $1.0e+1$ | $1.1e+1^*$ | $2.0e-1$ |
| | 5M_production | $4.0e-1$ | $3.9e+0^*$ | $3.8e+0^*$ | $3.9e+0^*$ | - |

Table 3: Comparison of the query throughput for different query scenarios, data sets and log stores (higher is better). Empty cells mark combinations which could not be executed successfully. Results within 10% of the best are marked with an asterisk.

For the *term(ID)* query scenario, the COPR store consistently outperforms all other implementations over all data sets, as can be seen in Table 3. The Scan store achieves a throughput of 0.44 queries per second for the *5M_production* data set. Considering the raw data size of 2 160 MB, this translates into a search throughput of 0.93 GB per second. In comparison, COPR reaches 3 300 queries per second. This translates into a search throughput of 6 961 GB per second, with a single thread. Such improvements are possible, because COPR allows us to avoid reading and decompressing the raw data in almost all cases. As an example, searching through 1PB of log data would require roughly 15 CPU cores with COPR to achieve a query latency of 10 seconds. Using a linear scan approach, this query latency would theoretically require 112 750 CPU cores.

COPR and CSC are both able to reach very low error rates for this query scenario. In [45] the error rate is defined as the number of incorrect classifications over the total number of samples. Since sketches and indices need to find the compressed data batches which contain relevant data, the error rate is defined as the number of found batches which do not contain the queried term, divided by the total number of batches. In other words, we measure the fraction of the overall data which is decompressed without contributing to the result. For the 5M_production data set, COPR achieves an error rate of $6.1e-7$ and CSC has a error rate of $1.9e-5$. To improve CSC's accuracy and overall query throughput, we additionally query for each n-gram within the query term and form the intersection of the individual results. Without this intersection, CSC would have a considerably higher error rate when querying for a single term.

Lucene does not have any false positives and is still slower than the COPR store. This suggests that the small error rate of the COPR store is outweighed by a more efficient evaluation compared to Lucene. LogGrep is at least 4 times slower than our linear scan implementation for the synthetic data sets. In addition, LogGrep could not successfully execute any query scenarios on the production data sets. However, we argue that the tests on synthetic data sets sufficiently show that LogGrep's approach of evaluating queries on the compressed data representations is unfeasible for our heterogeneous data sets and query use cases.

While term queries represent the intended and ideal use-case for token based indexing structures, contains searches within or across token boundaries are an important, necessary use-case. For the *contains(ID)* scenario, we again search for randomly generated sequences of 16 English letters. The sketch based COPR and CSC stores need to split these tokens into their n-grams and use these to find matching data batches. Lucene can rely on a linear scan of its token dictionary to support these queries. LogGrep and the Scan store always perform sub-string searches, since they have no concept of tokens. Figure 6 compares the query throughput of all relevant log store and data set combinations for this query scenario.

The COPR and CSC stores exceed Lucene's query throughput by at least two orders of magnitude for all data sets. Compared to the linear scan performance, Lucene only achieves a speed-up factor between 4.1 and 13.6, while COPR is between 708 and 3 456 times faster. For this query scenario, COPR and CSC both achieve similar error rates below $6.1e-4$. Lucene's dictionary scan needs to read and parse through the token dictionary for each query execution. This appears to be significantly more expensive than the n-gram lookups performed within the sketches and the low number of unnecessary batch decompressions.

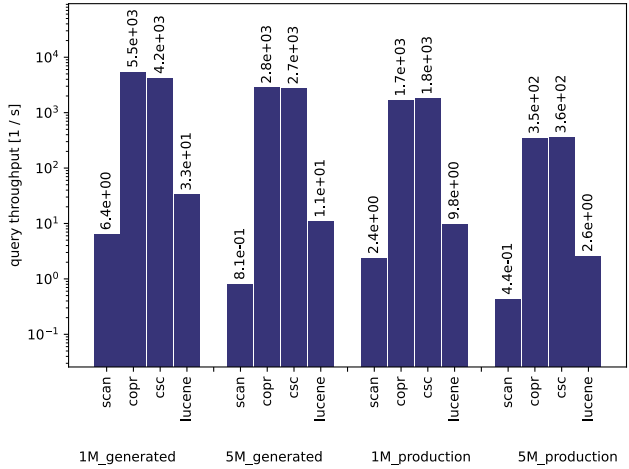

Figure 6: Throughput comparison for *contains(ID)* queries (higher is better).

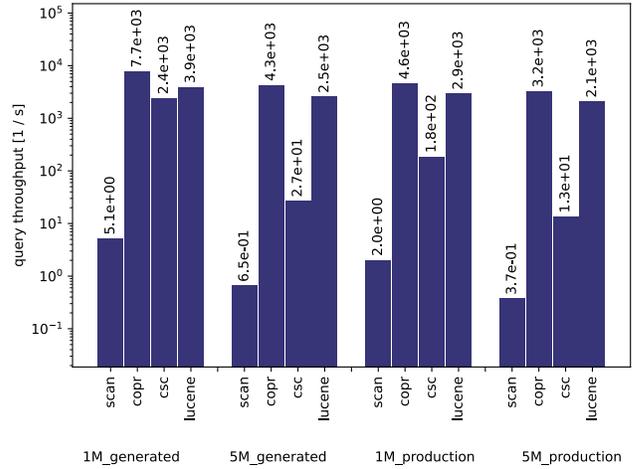

Figure 7: Throughput for *term(IP)* queries (higher is better).

The *IP* query scenarios search for log entries with a randomly generated, partial IP address (e.g., "192.130.100"). The main difference to the *ID* queries is the selectivity of the term's included n-grams. While tri-grams of randomly generated English letters still have a very high selectivity, most combinations of three numbers appear within a large fraction of the data batches. Therefore, the CSC sketch cannot profit from the n-grams to improve its error rate for the *term(IP)* query scenario. Figure 7 shows the query throughput of all relevant log store and data set combinations for these queries. As we can see, the COPR store achieves a throughput up to two orders of magnitude higher than the CSC store. For the 5M_production data set, COPR has an error rate of $1.2e-6$. In comparison, the CSC sketch has an error rate of $2.0e-2$. While COPR is able to maintain extremely low error rates even for individual tokens, the CSC sketch needs to intersect the results of multiple, highly selective tokens to reach similar error rates. The error rate of the CSC sketch can be improved to $2.0e-3$ with a second repetition within the sketch. However, this would double the required ingest time and slow down all other query scenarios because of the higher sketch evaluation cost.

If sketches have to rely on n-grams with a low selectivity for contains queries, they can only offer slight improvements in query speed over a linear scan. In this case, the dictionary scan of an inverted index can offer a higher throughput than a sketch based approach with indexed n-grams. This can be seen for the *contains(IP)* query scenario in Table 3.

As our last experiment, we look at *term(extracted)* queries which search for query terms taken from the data sets. These query terms appear on average in 0.4% of all batches for the 1M_generated data set, in 1.7% for the 1M_production data set, in 2.9% for the 5M_generated data set and in 4.8% for the 5M_production data set. Table 3 shows similar improvements for COPR, CSC and Lucene. Since the queried terms already match a relevant fraction of the data, the query throughput is mainly constrained by the processing of the matching batches. The index and sketch evaluation overhead, as well as a comparatively low number of false positive matches, only have a limited influence on the query throughput in this scenario. Still, COPR achieves the highest throughput across all four data sets, due to its efficient evaluation and high accuracy.

## 6 PRODUCTION EVALUATION

Dynatrace Grail performs detailed, internal self-monitoring regarding query executions and data storage. We can use this self-monitoring data to evaluate the impact of our COPR sketch within production systems. First, we will take a look at two exemplary, anonymous customer queries from our production systems. Both queries are formulated in the Dynatrace Query Language (DQL) [8]. The first query, shown below, searches for logs with a specific value for the field *userapp*, extracts some information from the content field via a parse command, selects certain log fields to be included in the result and sorts by time. Overall, it has to search through 34 billion log records, spread across 12 030 segments. However, since the provided *userapp* filter is highly selective, COPR can avoid the access to almost all potentially relevant data. Only two records were loaded from disk, allowing the query to search through 3 596 GB per second and CPU core. This scan rate is measured in relation to the uncompressed size of the ingested data. We further consider this example to serve as a testament to the high accuracy of the COPR membership sketch, since a false-positive rate of just 0.1% would already lead to roughly 34 million records being loaded from disk.

```
fetch logs
| filter matchesValue(userapp, "<term>")
| parse content, "<pattern>"
| fields timestamp, <fields>
| sort timestamp, direction: "<dir>"
```

The second query, from a different customer, simply counts the number of logs per log level (debug, info, warning, etc.) over the last 90 days. As this query does not exclude any log records through filters, the execution engine has to access and decompress all of the customers' 2.2 billion log records, spread across 1 159 segments.

In this extreme case, the scan rate drops to 2.5 GB per second and CPU core, over 1 400 times slower than the first query.

```
fetch logs, from:now() – 90d, to:now()
| summarize count = count(), by:{loglevel}
| sort count
```

An internal study of all executed customer queries confirms the same effects shown by our examples. Queries with highly selective filters, matching around one millionth of the data, achieve scan rates of up to 4 200 GB per second and CPU core. This rate is also remarkably close to the benchmark results achieved for the *term(ID)* query in Section 5, which represents a best-case scenario. When the filters match around 1% of the data, the scan rate drops to roughly 90 GB per second and CPU core. This also shows how severe the performance impact of membership sketches with higher false-positive rates, for example 1%, would be in practice. For queries accessing almost all records, a scan rate of up to 2.5 GB per second and CPU core is achieved, the same value observed in the second example query. This result also roughly represents the query performance expected without the search space reduction achieved through COPR, supporting our earlier argument that systems relying exclusively on highly parallel query execution cannot provide low query latencies and low analysis costs at the same time.

Across our production systems, Dynatrace Grail stores on average 2.1 GB of ingested data, corresponding to 1.4 million log records, within a single segment. A single log record on average consists of 1.5 KB of data in total and typically has several dozen key-value pairs, called fields, in addition to the log line itself. These fields add further context information to the records, e.g., identifying the machine or process from which the log was collected. All of these fields are indexed within COPR to support efficient filters on them. Even so, COPR only introduces a storage overhead of 1.1% of the ingested data size on disk.

## 7 CONCLUSION

In this work we have introduced a novel probabilistic membership algorithm called COPR that allows to answer multi-set multi-membership queries. We have evaluated this new structure compared to existing inverted indexers and membership sketches in terms of generation overhead, query throughput, and false positive rate. In our experiments COPR managed to outperform existing algorithms in terms of space overhead (e.g., requiring 90% less than Lucene) while also improving query speed compared to other sketches and indexing algorithms used in industrial settings for most scenarios and keeping an extremely low false positive rate.

The structure offers both a mutable version as well as an immutable version. The mutable structure efficiently adds new items via the secondary lookup map and allows usage on live data. While this mutable structure is efficient enough to be used in practice our algorithm additionally offers to transform the sketch into an immutable structure to reduce the memory footprint to a minimum for completed batches of data. The combination of the two types allows the use of COPR in practical, industrial applications.